\documentstyle[12pt,twoside,fleqn,epsf,espcrc2]{article}

\newcommand{\be}{\begin{eqnarray}}
\newcommand{\ee}{\end{eqnarray}}

\newcommand{\AmS}{{\protect\the\textfont2
  A\kern-.1667em\lower.5ex\hbox{M}\kern-.125emS}}

\title{Wee partons in large nuclei: from virtual dream to hard reality
\thanks{Plenary talk at Quark Matter '95, Monterey, CA.}}

\author{Raju Venugopalan \address{Institute for Nuclear Theory,
        University of Washington,\\
        Seattle, WA 98195 USA}}

\begin{document}
\maketitle

\begin{abstract}
We construct a weak coupling, many body theory to compute parton distributions
in large nuclei for $x\ll A^{-1/3}$. The wee partons are highly coherent,
non--Abelian Weizs\"{a}cker--Williams fields. Radiative corrections to the
classical results are discussed.  The parton distributions for a single nucleus
provide the initial conditions for the dynamical evolution of matter formed in
ultrarelativistic nuclear collisions.

\end{abstract}

\section{Introduction}

The title of this talk refers to wee partons in large nuclei. The first part
of this section will be a brief discussion of the theory and phenomenology of
wee parton distributions in QCD. The latter part of this section (and
subsequent sections) will be about wee partons distributions in large nuclei
and their relevance to experiments at RHIC and LHC.

\subsection{A brief introduction to the low $x$ problem in QCD}
\vspace*{0.4cm}

One of the more interesting problems in QCD today is to understand the
space--time (or equivalently, momentum) distributions of ``wee partons" in
hadrons and nuclei. The phrase ``wee parton" was coined by Richard Feynman~
\cite{Feynman} to describe those partons which carry a very small fraction of
the longitudinal momentum of the parent hadron or nucleus. This momentum
fraction is the Feynman $x$ variable and for wee partons, $x\ll 1$. To lowest
order in the coupling constant, the Feynman $x$ variable is equal to the
well known Bjorken $x$ of deeply inelastic scattering experiments.

Why is this problem of wee parton distributions interesting? To understand this
let us first briefly review the usual
Dokshitzer--Gribov--Lipatov--Altarelli--Parisi branching process in
perturbative QCD~\cite{Kwiecinski}. This branching process describes a virtual,
ladder--like cascade, where partons with higher values of transverse momentum
$k_t$ and $x$, ``split" into softer partons with lower values of $k_t$ and $x$
and so on along the ladder. Each rung in the ladder provides logarithmic
contributions to the cross section of the sort
\be
\alpha_S \int {d^2 k_t\over k_t^2} \int {dx\over x} \, .
\ee
In the Double Leading Logarithmic Approximation (DLLA), both $x$ and $k_t$ are
ordered along the ladder; this is equivalent to summing leading logs in both
variables. In the DLLA, the gluon distribution function, for $x\ll 1$, grows
as~\cite{Gross}
\be
x {dN\over dx} (x,Q^2; Q_0^2) \sim
exp\left(2\sqrt{\log(1/x)\log\log(Q^2/Q_0^2)}
\right)
\, ,
\ee
where $Q_0^2$ and $Q^2$ are the respectively the lowest and highest values of
$k_t^2$ along the rungs of the ladder. The latter value is usually identified
with the momentum transfer squared in deeply inelastic scattering experiments
while the former typically sets the factorization scale.

However, if $x$ is very small and $Q^2$ is not large enough, the leading
logarithmic terms in $x$ will dominate and it may be appropriate to sum those
alone. In this approximation, known as the BFKL approximation~\cite{BFKL}
(after the initials of the authors), one has ordering along the ladder rungs
only in $x$. In the Regge--Mueller language, this corresponds to a t--channel
exchange of a ``perturbative pomeron" in the virtual photon$+$ target
scattering process. The gluon distribution then has the form
\be
x {dN\over dx}(x,Q^2; Q_0^2)  \sim \sqrt{{Q^2\over Q_0^2}} \,\,
{1\over x^{\lambda}} \, ,
\ee
where $\lambda=4\alpha_s\log(2)$. This behaviour of the gluon distributions is
often called the Lipatov enhancement. An interesting feature about the BFKL
approach is that, unlike the DLLA,  it is {\it not} a twist expansion (as in
the Operator Product Expansion).

Regardless of whether the DLLA or the BFKL approximations are more appropriate
for the $x$ values of interest, they have one outstanding feature in common:
for $x\ll 1$, the density of wee partons ($dN/dx$) grows rapidly. In both
cases, it grows faster than $1/x$. At sufficiently small values of $x$, one
enters a regime where partons from neighbouring ladders overlap spatially. This
regime of ``high density" QCD~\cite{Levin} is of interest because it describes
the many body behaviour of the quanta of the fundamental theory.

One consequence of the overcrowding of partons is that two soft partons may
recombine to form a harder parton. Another consequence of the overcrowding is
the screening of parton--parton interactions by the cloud of surrounding wee
partons~\cite{MueQiu}. Both of these processes may inhibit the growth of parton
distributions and cause them to saturate at some critical value of $x$.
Incidentally, the saturation of parton distributions as $x\rightarrow 0$ is
required by unitarity--the so--called Froissart bound. Because these many body
phenomena are very complex, it is especially gratifying to realize that they
may (to a degree) be understood using well known, weak coupling techniques~
\cite{Mueller}.

Why are weak coupling techniques applicable in this problem? The only scale in
the problem at low $x$ is $\rho$, the density of partons. The coupling constant
$\alpha_S$ will run as a function of this scale and if $\Lambda_{QCD}\ll \rho$,
then $\alpha_S(\rho)\ll 1$. {\it This qualitative argument can be proven
rigorously and is the basis of all that follows}.

Thus far, we have focused on some of the theoretical reasons why one must
better understand wee parton distributions. These theoretical impulses have
been around from the pre--history of QCD. What is re--generating interest in a
previously moribund field is a new generation of experiments, which have begun
to probe the high density phase of QCD. Foremost among these is HERA, the
electron--proton collider experiment at DESY~\cite{ZEUS}. At HERA, $x$'s as low
as $x\sim 10^{-5}$ and $Q^2$ as high as $Q^2\sim 10^5$ GeV$^2$ are attained.
One of the observables measured at HERA is $F_2^{ep} (x,Q^2)$, the
electromagnetic form factor of the proton. At the above mentioned values of
$x$, $F_2$ continues to rise with decreasing $x$, for a wide range of $Q^2$
values.

Since $F_2$ can be related directly to the sea quark distributions and
indirectly to the gluon distributions, these experimental results seem to
confirm the behaviour predicted by the theoretical models discussed above. The
data are apparently not good enough to distinguish between the DLLA and BFKL
scenarios--this issue is controversial and will hopefully be settled
conclusively in the near future.

Another class of experiments which probe the region of low $x$ are the deeply
inelastic scattering experiments at Fermilab (E665) and CERN (NMC).
These experiments are fixed target experiments where a high energy lepton
scatters off a nuclear target. It is observed~\cite{Melanson} that the ratio
$F_2^{A}/{A F_2^N}$ shows a significant depletion at $x\leq 10^{-2}$--
a phenomenon known as ``shadowing". The shadowing of nuclear distributions at
low $x$ shows an unambiguous $A$ dependence of $A^{0.9}$. Can shadowing be
understood as a consequence of the faster (i.e., at higher $x$) saturation of
nuclear distributions as opposed to hadron distributions? Naively, the answer
is
yes, but a conclusive answer will have to be a quantitative one.

Finally, there are the collider experiments at RHIC and LHC which will also
probe fairly low $x$'s ($\sim p_t/\sqrt{s})$. The wee parton distributions of
the nuclei will provide the initial conditions for the evolution of the system
into a quark gluon plasma. Further, most of the observables in a nuclear
collision will depend sensitively on the wee parton distributions. In the
following sub--section we will discuss briefly the problem of wee parton
distributions in large nuclei. Subsequent sections will address the problem in
greater detail.

\subsection{Wee partons in large nuclei: an outline}
\vspace*{0.4cm}

In recent papers~\cite{LV1,LV2}, Larry McLerran and I have argued that large
nuclei are an excellent theoretical tool to study the low $x$ problem. We show
that for large nuclei, in the range of $x\ll A^{-1/3}$, the large density of
partons introduces a scale $\mu^2\approx A^{1/3}$ fm$^{-2}$, which can be
identified as the average {\it valence quark} color charge squared per unit
area. When $\mu^2$ is large ($A$ is large) the coupling constant
$\alpha_S(\mu^2)$ is small and a weak coupling expansion can be employed to
compute quark and gluon distributions.

On the phenomenological side, we expect that this first principles calculation
of quark and gluon distributions in QCD will help clarify our understanding of
the initial stages of nuclear collisions at RHIC and LHC. There exist state of
the art partonic cascade models which provide an excellent initial orientation
to this problem~\cite{geiger}. However, since cascade models do not accurately
simulate the wee parton sea, the enhancement of the momentum scale due to the
large parton density will, in comparison, likely lead to i) higher
temperatures, ii) enhanced minijet production and iii) an enhanced contribution
of ``intrinsic" strangeness and charm, than those predicted in these models.

We will now discuss how the paper is organized. We formulate the problem
of low $x$ parton distributions as a many body problem in the presence of an
external source. The source here is a sheet of static, valence quarks localized
in the longitudinal direction but infinite in extent and uniformly varying in
the transverse direction. Next, we write down a partition function for this
system which includes a stochastic averaging over the external sources of color
charge. This procedure introduces an additional dimensionful parameter $\mu^2$
in the theory. It is defined as the average valence quark colour charge squared
per unit area.

The approach we adopt to solve the above mentioned problem is as follows: we
first obtain the classical background field in the presence of the source. It
turns out that there is a very simple solution to the classical field equations
in which the only non--zero fields are the transverse vector fields which are
zero ahead of the source and two dimensional pure gauge fields behind the
source. The gluon distribution functions in the background field can thus be
expressed as correlation functions of a two dimensional, Euclidean field
theory.

For parton transverse momenta $\alpha_s^2\mu^2 << k_t^2$, which we identify as
a weak coupling regime, the gluon distributions obey the well known
Weizs\"{a}cker--Williams distribution scaled by $\mu^2\sim A^{1/3}$. In the
strong coupling regime $k_t^2<<\alpha_s^2 \mu^2$, the distribution functions
still vary as $1/x$ but the transverse momentum dependence is changed. This
likely reflects the sharp decay of correlation functions in transverse space.
We have not succeeded in obtaining an analytical expression for the transverse
momentum dependence--but have succeeded instead in writing an algorithm to
solve the equations numerically using a Monte Carlo procedure.

We next outline the procedure to compute small fluctuation Green's functions
for scalars, vectors and fermions in this classical background field. The
detailed computations, performed in collaboration with A. Ayala, J. Jalilian--
Marian and L. McLerran are reported elsewhere~\cite{LV3,LV4,LV5}. These Green's
functions are useful when one computes the higher order contributions in
$\alpha_S$ to the quark and gluon distribution functions. The fermion Green's
functions are used to determine the sea quark distributions. We find that the
enhanced momentum scale arising from the higher density of partons leads to an
enhancement of the strange and charm quark contribution to the nuclear
wavefunction.

We then proceed to discuss briefly higher order corrections to our results
--specifically, the one loop corrections to the classical background field and
the origin of the Lipatov enhancement in our approach. We also speculate how
our solutions may be iterated to all orders in $\alpha_S$. Finally, we discuss
recent progress in extending our approach to nuclear collisions and end our
presentation with a brief summary.

\section{Formulation of the problem}
\vspace*{0.4cm}

In this section we will formulate the problem of calculating the distributions
of partons in the nuclear wavefunction as a well defined many body problem.
To
do this we will work in the Infinite Momentum frame and use the technique of
Light Cone quantization. We will also be working in Light Cone gauge $A^+=0$.
There are several advantages to these choices. One of these is the fact that it
is only in Light Cone gauge that partons have a manifest physical
interpretation as the quanta of the theory. Further, if we work on the Light
Cone, it is possible to construct a simple and intuitive Fock space
basis~\cite{Brodsky} on which our formalism relies heavily. One consequence is
that  the electromagnetic form factor of the hadron $F_2$, which is measured in
deeply inelastic scattering experiments, is simply related to parton
distributions by the relation~\cite{dkta}
\be
F_2(x,Q^2)=\langle \int^{Q^2} d^2 k_t dk^+ x \delta(x-{k^+\over P^+})\sum_{
\lambda=\pm} a_{\lambda}^{\dagger} a_\lambda\rangle \, .
\ee
In the above, $P^+$ is the momentum of the nucleus, $k^+ = (k+k_z)/\sqrt{2}$
and $k_t$ are the parton ``longitudinal momentum" and transverse momenta
respectively, $Q^2$ is the momentum transfer squared from the projectile and
$a^{\dagger} a$ is the number density of partons in momentum space. Hence, in
principle one only needs up to integrate the calculated distributions to the
scale $Q^2$ to make comparison with experiment. One difficulty in the Light
Cone formalism is the problem of choosing appropriate boundary conditions. This
issue is beyond the scope of this talk. For an alternative approach to the
problem, the reader may consult the recent paper by Makhlin~\cite{Makhlin}.

\subsection{A static source on the Light Cone}
\vspace*{0.4cm}

We begin by making  several physically plausible assumptions. First,
if the density of partons per unit area at low $x$ is large, and weak coupling
techniques are applicable,  the recoil experienced by the valence quarks due to
bremsstrahlunging of low $x$ partons is small--the valence quarks obey straight
line trajectories. We can thus replace the valence quarks in the infinite
momentum frame nuclear wavefunction by static, external sources of charge.
Next, if we require that the nucleus be Lorentz contracted to a
size which is much smaller than the wavelength of the parton in a frame
co--moving with its longitudinal momentum, then
\be
2 Rm/P<<1/xP \Longrightarrow x<<1/Rm \, ,
\ee
where $R$ is the nuclear radius, $P$ the momentum of the nucleus and $m$ the
nucleon mass. From the above inequality, we deduce that when $x<<A^{-1/3}$, the
partons see a sheet of static color charges of large transverse extent but
localized in the longitudinal direction. Since the average transverse momentum
scale $k_{\perp}>>1/R$, the valence quark distribution can be taken to be
uniform and infinite in extent in the transverse direction.

It is appropriate to use Light Cone co-ordinates to investigate the dynamics of
the infinite momentum frame wavefunction of the nucleus. The conventions used
here are discussed in Ref.~\cite{LV1}. In the Light Cone formalism, the
``external" current experienced by the low $x$ partons due to the static,
valence quark charges takes the form
\be
J^{\mu}_a = \delta^{\mu +}\rho_{a}(x^+,\vec{x}_{\perp}) \delta(x^-) \, .
\label{extcur}
\ee
In the above, $x^{\pm}=(t \pm x)/\sqrt{2}$, where $x^+$ is the Light Cone time.

The Light Cone Hamiltonian $P^-$ ( which generates translations in $x^{+}$), in
the presence of an external source, can be written as
\be
	P^- & = & { 1 \over 4} F_t^2
+ {1 \over 2} \left( \rho_F + D_t \cdot E_t
\right) {1 \over {P^{+}}^2} \left( \rho_F + D_t \cdot E_t
\right) \nonumber \\
   &+&{1\over 2}\psi^\dagger (M-\not \! P_t ) { 1 \over P^+ } (M+\not \! P_t)
\psi \, .
\ee
Here $\rho_F$ is the charge density due to the external source plus the
dynamical quarks, $A_{t}$ and $\psi$ are the dynamical vector and spinor
fields, $E_t = \partial_{-} A_t$ is the transverse electric field and $F_t^2$
is the transverse field strength tensor. In writing the above expression, we
have made use of the constraint equations on the Light Cone to eliminate the
non--dynamical fields.

\subsection{Ground state expectation values in the presence of
external sources}

\vspace*{0.4cm}

In QED, the infinite momentum frame wavefunction of the system with the
external source in Eq.~(\ref{extcur}) is a coherent state~\cite{LV1}. We have
not succeeded in doing the same in QCD and have concentrated instead on
computing ground state expectation values in the presence of external sources.
The partition function for the ground state of the low $x$ partons in the
presence of the valence quark external source is
\be
Z=<0|e^{iTP^-}|0>=\lim_{T\rightarrow i\infty} \sum_{N} <N|e^{iTP^-}|N>_{Q} \, .
\label{qmsum}
\ee
The sum above also includes a sum over the color labels of the source of color
charge (denoted by $Q$) generated by the valence quarks.

In principle, for quantized sources of color charge, evaluating the trace over
different values of the color charge is difficult. However, for a large
nucleus,
the problem can be simplified considerably. Visualize the transverse space as
a grid of boxes of size $d^2 x_t >> 1/ \rho_{valence}\sim A^{1/3}$ or
equivalently, of parton transverse momenta $q_t^2<< A^{1/3}$ MeV$^2$. In this
kinematic range, the number of valence quarks, and therefore the charge $Q$ in
each box, will be much greater than one. A large number of charges corresponds
to a higher dimensional representation of the color algebra--the sum of the
color charges of the valence quarks in a grid can be treated classically.

Further, if the total charge in the grid is much less than the maximum possible
charge the central limit theorem tells us that the density of states
corresponding to charge $Q$ is a Gaussian $e^{-Q^2/2\mu^2}$. Summing over the
color labels of the states is therefore equivalent to introducing in the path
integral the measure
\be
\int [d\rho] \exp\bigg(-{1\over {2\mu^2}}\int d^2 x_t \rho^2 (x_t) \bigg) \, .
\label{measure}
\ee
This ensures that
\be
<\rho^a (x_t)\rho^b (y_t)>= \mu^2 \delta^{ab}\delta^{(2)}(\vec{x_t}-\vec{y_t})
\, ; \, <\rho^{a} (x_t)> = 0 \, .
\ee
Above, the variance $\mu^2$ is the average valence quark color charge squared
per unit area. It can be written as
\be
\mu^2 = \rho_{valence} <Q^2> \equiv {3A\over {\pi R^2}} {4\over 3} g^2 \sim
1.1 A^{1/3} fm^{-2} \, ,
\ee
where $<Q^2>=4 g^2/3$ is the average charge squared of a quark.

We can now write the path integral representation of the partition function $Z$
in the Light Cone gauge $A_{-}=0$ as
\be
	Z & = & \int~ [dA_t dA_+] [d\psi^\dagger d\psi] [d\rho]
 \nonumber \\
& & exp\left( iS +ig\int d^4x A_+(x)\delta (x^-)
\rho (x)  - {1 \over {2\mu^2}} \int d^2x_t \rho^2 (0,x_t) \right) \, .
\label{funp}
\ee
Hence, the result of our manipulations is to introduce a dimensionful parameter
$\mu^2 \approx 1.1 A^{1/3}$ fm$^{-2}$ in the theory.  If we impose the current
conservation $D_\mu J^\mu =0$ and integrate over the external sources $\rho$,
we obtain an expression~\cite{LV1} containing modified propagators and
vertices.

We shall however attempt to solve the above many body problem in the following
fashion: we first solve for the classical background field, compute the small
fluctuation determinant in the background field to obtain Green's  functions
and propagators, use these to compute the first radiative corrections to the
background field and systematically iterate the process to all orders.

\rm\baselineskip=14pt
\section{The Classical Background Field}
\vspace*{0.4cm}

We first discuss the problem of computing the solution of the classical
equations of motion for the gluon field in the presence of a source which is a
delta function along the Light Cone. The second part of this discussion is
about computing correlation funtions in this background field. Notice from
Eq.~(\ref{funp}) that there is still an integral over $\rho$ with a Gaussian
weight to average over at the end. At each order in $\alpha_S$, this procedure
is equivalent to summing the theory to all orders in the effective coupling
$\alpha_S \mu/k_t$.

\subsection{Solutions to the Classical Equations}
\vspace*{0.4cm}

The equations of motion are
\be
	D_\mu F^{\mu \nu} = g J^\nu \,,
\ee
where $J$, the classical Light Cone source, may be represented as
\be
	J^\mu_a = \delta^{+\mu} Q_a(x^+,x_t) \delta(x^-) \, .
\label{source}
\ee
We will work in Light Cone gauge where $A_- = - A^+ = 0$.

There exists a solution of the equations of motion for this problem, where
the longitudinal component $A_+$, which is not zero by a gauge, vanishes by the
equations of motion
\be
	A_+ = -A^- = 0 \, .
\label{conf}
\ee
The only non--zero components of the field strength are the
transverse components which we require to be of the form
\be
	A_i(x) = \theta (x^-) \alpha_i (x_t) \, .
\ee
This functional form is what we expect for a classical field generated by a
source traveling close to the speed of light with $x = t$. For $x > t$, the
source has not yet arrived, and for $x < t$, the source should produce a field.

If we further require that $F^{ij} = 0$ (where i and j are transverse
components), we see that we have a solution of the equations of motion as long
as
\be
	\nabla \cdot \alpha = g \rho(x_t) \, .
\label{gauge}
\ee
Here $\rho $ is the surface charge density associated with the current $J$.
There is no dependence on $x^-$ because we have factored out the delta
function.  The dependence on $x^+$ goes away because of the extended current
conservation law $D_{\mu}J^{\mu}= 0$ in the background field.

The condition that $F^{ij} = 0$ is precisely the condition that the field
$\alpha $ is a gauge transform of the vacuum field configuration for a two
dimensional gauge theory. The field configuration which is a gauge transform of
the vacuum field configuration for a two dimensional field theory may be
written as
\be
	\tau \cdot \alpha_i = -{1 \over ig} U \nabla_i U^\dagger\, .
\ee
The dependence of the $U$ fields on the surface charge density $\rho$ is then
given by the relation
\be
	\nabla \cdot U \nabla U^\dagger = -ig^2 \rho(x_t) \, .
\label{ugauge}
\ee

We have not been able to construct analytical solutions for the above equation
for arbitrary dependence of the surface charge density on $x_t$. However,
recently we have found a numerical algorithm which will solve the above
equation on a 2--D lattice using a Monte Carlo procedure~\cite{Matthias}.

\subsection{Computing Correlation Functions}
\vspace*{0.4cm}

To compute correlation functions associated with our classical solutions,  we
must integrate over  all color orientations of the external sheet of charge.
This is equivalent to computing the expectation value of $\langle\alpha_i(x_t)
\alpha_j(y_t)\rangle $ with the measure (expressed in terms of the compact
fields $U$)
\begin{eqnarray}
	\int [dU] ~\exp\left( -{1 \over {g^4\mu^2}}\int d^2x_t {\rm Tr}
\left[\vec{\nabla} \cdot (U{1 \over i}\vec{\nabla} U^\dagger ) \right]^2
\right)
{\rm det}(\vec{\nabla} \cdot \vec{D} )
\label{fluct}
\end{eqnarray}

We see that the measure for this theory is that for a two dimensional Euclidean
field theory where the expansion parameter is $\alpha_S^2 \mu^2/k_t^2$. Since
the theory is two dimensional, it should be straightforward to compute
correlation functions to all orders in $\alpha_S^2 \mu^2/k_t^2$ using lattice
Monte Carlo methods. Further, since the theory is ultraviolet finite, there
should be no problems extrapolating to the continuum limit.

Since we can solve Eq.~(\ref{ugauge}) numerically, the correlation functions
can still be expressed as an integral over the Gaussian measure in $\rho$. We
then do not need to worry about the nasty Fadeev--Popov term in the above
measure.

The correlation function for the computation of the transverse momentum
dependence of the structure function may be simply evaluated to be
\be
	D(k_t) = {1 \over \alpha_S} H(k_t)
\ee
where $D$ is the propagator for the two dimensional theory
\be
	(2\pi )^2 \delta^2 (k_t-q_t) \delta_{ij} D(k_t) = <\alpha_i \alpha_j>
\ee

The relation between distribution functions and propagators is straightforward
and is discussed explicitly in Ref.~(13). For instance, for a scalar field
$\phi^{\alpha}(x^+,x_t)$, in the fundamental representation, the distribution
function is given by the relation
\be
	{{dN} \over {d^3k}} =  2 i~{(2 k^+)  \over {(2\pi)^3}}~
\sum_\alpha D^{\alpha \alpha }
(x^+,\vec{k}, x^+,\vec{k} ) \, .
\label{disgre}
\ee
The propagator in the above equation is defined as
\be
	D(x^+,\vec{k}, y^+,\vec{q} ) = \int d^3x d^3y ~e^{-ikx+iqy} G(x,y) \,
,
\ee
where
\be
	G^{\alpha \beta } (x,y) =  -i< \phi^\alpha (x) \overline
\phi^{~\beta} (y) >\, .
\ee

Now, in the correlation functions of the classical field $\langle A_t (x_t) A_t
(y_t)\rangle $, the dependence on $x^-$ is only through a step function, and
upon Fourier transforming gives only a factor of $1/k^+$. The distribution
functions associated with this field, to all orders in $\alpha_S^2 \mu^2$, have
the general form~\cite{LV2}
\be
	{1 \over {\pi R^2}} {{dN} \over {dx d^2k_t}} =
{{(N_c^2-1)} \over \pi^2} {1 \over x}~
{1 \over \alpha_S}H(k_t^2/\alpha_S^2 \mu^2)\, .
\ee

When the relevant momentum scale is $k_t^2 >> \alpha_S^2 \mu^2 $, the theory is
in the weak coupling region and may be evaluated perturbatively.  When
$y\rightarrow \infty$,
\be
H(y)= 1/y \, ,
\ee
and we obtain for the distribution function the Weizs\"{a}cker--Williams result
\be
	{1 \over {\pi R^2}} {{dN} \over {dxd^2q_t}} = {{\alpha_S \mu^2(N_c^2-1)}
\over \pi^2} {1 \over {xq_t^2}}\, ,
\label{weiz}
\ee
scaled by $\mu^2$.

For small values of $k_t^2<<\alpha_S^2 \mu^2$, we are in the strong coupling
phase of the theory.  In this phase of the theory, we expect that there should
be no long range order. Correlation functions of $x_t$ should die exponentially
at large distances, or alternatively the Fourier transform of correlation
functions should go like
\be
{dN\over {d^2 k_t}} \sim {1\over {k_t^2 + \alpha_S \mu^2}} \, ,
\label{StCple}
\ee
which plateaus off to a constant for small momentum. This is shown in Fig.~1.
Hence, $\alpha_S\mu$ is like a Debye scale which  guarantees the finiteness of
the gluon distribution functions for small momentum.

\begin{figure}[htb]
\begin{minipage}[t]{80mm}
\framebox[75mm]{\rule[-26mm]{0mm}{52mm}}
\epsfbox[30 -5 0 30]{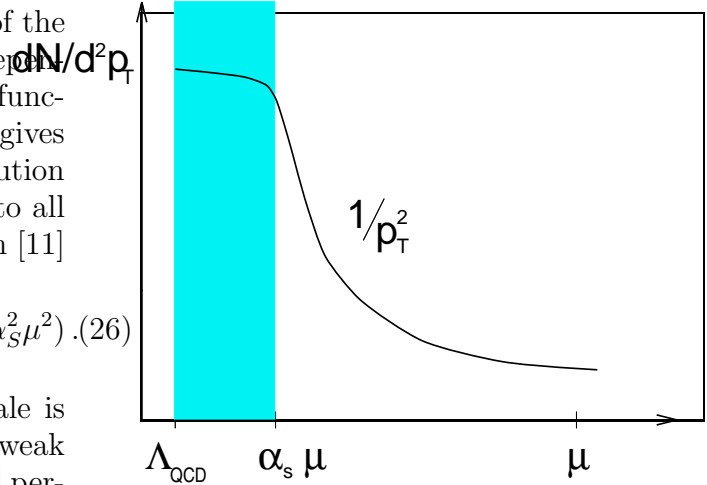}
\caption{Transverse momentum distribution for Weizs\"{a}cker--Williams fields}
\end{minipage}
\end{figure}

Note that the gluon distribution function in the weak coupling
Weizs\"{a}cker--Williams regime is additive in $A$. If the behaviour in the
strong coupling regime would resemble that in Eq.~(\ref{StCple}), the gluon
distribution function would then be proportional to $A^{2/3}$. If we integrate
over all $k_t^2$ to some scale $Q^2$, the integrated distribution would
probably have a dependence which lies between $A^{2/3}$ and $A$. One must keep
in mind however, the fact that higher orders will likely modify the $A$
dependence of the classical distribution functions.

\section{Small fluctuations and higher orders in perturbation theory}
\vspace*{0.4cm}

In previous sections, we discussed only the {\it classical} solutions of the
Yang--Mills equations in the presence of the valence quark source defined in
Eq.~(\ref{source}). In this section, we will outline a procedure to
systematically compute quantum corrections to our background field to all
orders. This procedure is exactly equivalent to the familiar Dyson--Schwinger
expansion in many body physics~\cite{Baym}.

We begin by considering small fluctuations around our classical background,
\be
A_{cl}=A_{cl}^0 + \delta A \, .
\label{fluct}
\ee
Substituting this in the partition function in Eq.~(\ref{funp}), we keep
only terms upto O($\delta A^2$) in the action. The small fluctuations
propagator
may be computed directly from the action or equivalently by substituting
Eq.~(\ref{fluct}) in the Yang--Mills equations, keeping terms linear in
$\delta A$, and solving the resulting eigenvalue equation.

The gluon Green's function is then obtained from the general expression
\be
G(x,y) = \sum_{\lambda} {{\delta A_{\lambda}(x)\delta A_{\lambda}^{*}(y)}\over
\lambda} \, .
\ee
In Ref.~(13), we obtained Green's functions for scalars, vectors and fermions
in
the non--Abelian Weizs\"{a}cker--Williams (NAWW) background field. It turns out
that the gluon Green's function we computed there in $A^+=0$ gauge, is
incorrect and is actually the Green's function in $A^-=0$ gauge instead. The
direct computation of the Green's function in Light Cone gauge $A^+=0$, is
hampered by the presence of the singular source term in the small fluctuation
equations of motion. The correct procedure for computing the Light Cone gauge
Green's functions is discussed in Ref.~(14). The reader is referred to
the papers Ref.~(13) and Ref.~(14) for relevant details.

The general expression for the two point function to all orders in the coupling
constant is given by
\be
\langle A_t A_t \rangle = \langle A_t^{cl}\rangle \langle A_t^{cl}\rangle +
\langle A_t^{qu} A_t^{qu}\rangle \, ,
\label{dyson}
\ee
where $<A_t^{cl}>$ is the classical field to all orders and $<A_t^{qu}
A_t^{qu}>$ is the small fluctuation Green's function in the background field.
The above expectation values also include a stochastic averaging over the
sources of external charge with the Gaussian weight discussed in
Eq.~(\ref{measure}). Recall that as a consequence,  there are two expansion
parameters in the theory: $\alpha_S$ and $\alpha_S \mu$. So one should be wary
of naive $\alpha_S$ counting!

The zeroth order contribution to $<A_t^{cl}>$ is the classical background field
we discussed in previous sections and is of order O($1/g$)$\sum_{1}^n (
\alpha_S \mu)^n$. The lowest order contribution to the distribution functions
then gives
\be
A_t^{(0)}A_t^{(0)}\propto {1\over {g^2}} (\alpha_S \mu)^{2n}
\longrightarrow  \alpha_S \mu^2  (n=1)\, ,
\ee
which is the result obtained in Eq.~(\ref{weiz}).

The next order contribution O($g^0$) to the distribution function has two
components. The first is from the small fluctuation Green's function
$\langle A_t^{qu} A_t^{qu}\rangle$ in Eq.~(\ref{dyson}) above. Recall that
the Green's functions we compute are directly related to distribution
functions by Eq.~(\ref{disgre}). In momentum space this is equivalent to both
legs of the distribution function having the same momentum. Performing the
computation, we find a term proportional to $\log(x)/x$. If we iterate this
procedure to all orders (the distribution functions now forming a ladder)--
summing the leading logs in $x$, we might get a series which looks like
\be
{1\over x}\left(1 + C\alpha_S \log x +{C^2\alpha_S^2\over 2}\log^2 x +\cdots
\right) \equiv {1\over {x^{1+C\alpha_S}}} \, .
\ee
The crucial point is: what is the coefficient $C$? How does it depend on
$\mu^2$? This still remains to be settled: we are working on it!

The other O($g^0$) contribution to the distribution function comes from the
term $A_t^{(0)}A_t^{(1)}$ in Eq.~(\ref{dyson}). Here $A_t^{(1)}$ includes the
one loop radiative correction to the background field. One can show~\cite{LV5}
from the operator equations of motion that $A_t^{(1)}$ satisfies the equation
\be
\partial_{-}^2 A^{-,a} + D_i (\partial_{-} A_i^{(1),a}) +
gf_{abc}\lim_{x\rightarrow y}\partial_{y^-} G_{ii}^{bc}(x,y)=J_{cl}^{+,a} \, .
\ee
The final term on the left hand side is the induced current in the background
field. We find that it has exactly the same structure as the classical current
$J_{cl}^+$ and its only effect is to renormalize the external charge
$\rho(x_t)$. It therefore also follows that $A_t^{(1)}$ has exactly the same
structure as $A_t^{(0)}$. The only change is that
\be
\alpha_S \longrightarrow \alpha_S (\mu^2) \, .
\ee

{\it We may therefore conclude that the only effect of the classical field
$\langle A_t^{cl}\rangle$ is to replace the bare coupling constant in the
ladder sum by a coupling constant which runs as a function of $\mu^2$.}
This result is not entirely surprising~\cite{Pokorski} but it is nevertheless
interesting to see how it arises in our formalism.

\section{From virtual dream to hard reality}
\vspace*{0.4cm}

In the preceding sections, we investigated the properties of the NAWW
field of a single, large nucleus. This is relevant for deeply inelastic
scattering off nuclei at small $x$. To understand nuclear collisions, however,
we need to understand how the NAWW fields evolve after the collision. In
addition, we need to address the question of how the virtual partons go ``on
shell" and undergo hard scattering. The scattering is hard because most of the
wee partons are in the ``hard" kinematic region $\alpha_S\mu\ll k_t\ll \mu$. In
the first part of this section, we will discuss ideas of Larry McLerran and
collaborators at Minnesota on the {\em classical} problem of the evolution of
the NAWW fields. The latter part will be a brief discussion of ``onium--onium"
scattering as formulated by A.~H.~Mueller.

\subsection{The dynamical evolution of classical non--Abelian
Weizs\"{a}cker--Williams fields in nuclear collisions}
\vspace*{0.4cm}

Recently, A.~Kovner, L.~McLerran and H.~Weigert~\cite{KLW}  managed to solve
the {\it classical} problem of the evolution of these fields. They showed that
at late times, the equations for the evolution of the fields are linear.

Below, we outline the important results in their paper and refer the interested
reader to their paper for further details. The Yang-Mills equation for the two
source problem is
\begin{eqnarray}
D_\mu F^{\mu \nu} = J^\nu (x)
\label{YMcoll}
\end{eqnarray}
where
\begin{eqnarray}
	J^{\pm} & = & \delta (x^{\mp}) \rho_{1,2} (x_\perp)\nonumber \\
        J^i & = & 0.
\end{eqnarray}
In the above, $\rho_{1}(x_t)$ and $\rho_{2}(x_t)$  are the color charge
distributions of the valence quarks of the two nuclei.

Before the two nuclei collide (for times $t < 0$), the above equations of
motion are satisfied by the classical field configuration where $A^{\pm}=0$ and
\begin{eqnarray}
        A^i  =  \theta (x^-) \theta(-x^+) \alpha^i_1 (x_\perp) +
 \theta (x^+) \theta (-x^-) \alpha^i_2 (x_\perp)
\end{eqnarray}
The two dimensional vector potentials are pure gauges (as in the single nucleus
problem!) and for $t < 0$ solve
\begin{eqnarray}
	\nabla \cdot \alpha_{1,2} = g^2 \rho_{1,2}(x_\perp) \, .
\end{eqnarray}
The interesting aspect of this solution is that the classical field
configuration does not evolve in time for $t < 0$! This is a consequence of
the highly coherent character of the wee parton cloud in the nuclei.

The above solution for $t < 0$ is a fairly straightforward deduction from the
single nucleus case. What is very interesting is that the above mentioned
authors find a solution to the field equations for $t > 0$. It is given by
\begin{eqnarray}
	A^{\pm} & = & \pm x^{\pm} \alpha(\tau, x_\perp) \nonumber \\
        A^i & = & \alpha_3^i (\tau,x_\perp) \, ,
\end{eqnarray}
where $\tau = \sqrt{t^2 - z^2} = \sqrt{2x^+x^-}$. This solution only depends on
the longitudinal boost invariant variable $\tau$ and has no dependence on the
space-time rapidity variable $y = { 1 \over 2} \ln{x^+ \over x^-}$. That the
solution is independent of $y$ suggests that the parton distributions will be
boost invariant for all later times. This result therefore justifies Bjorken's
ansatz~\cite{bj1} for the subsequent hydrodynamic evolution of the system.

The equations that result for $\alpha(\tau,x_\perp)$ and
$\alpha_3(\tau,x_\perp)$ are highly non--linear. The detailed expressions are
given in Ref.(22).

Asymptotically, for large $\tau$,
\begin{eqnarray}
	\alpha(\tau, x_\perp) & \rightarrow &
	V(x_\perp) \epsilon(\tau, x_\perp) V^\dagger
	(x_\perp) \nonumber \\
        \alpha^i_3(\tau, x_\perp) & \rightarrow &
	V(x_\perp)\left[ \epsilon^i(\tau, x_\perp)
	 -{1 \over {ig}} \partial^i\right]
	V^\dagger (x_\perp)  \, ,
\end{eqnarray}
where the value of the gauge transformation $V(x_\perp)$ is determined by the
field equations.  It results from solving the non-linear time evolution
equations for the fields.

The equations of motion for the fields in the asymptotic region are linear for
$\epsilon$ and $\epsilon^i$. The solutions to these equations at asymptotically
large $\tau$
are of the form
\begin{eqnarray}
	\alpha^a(\tau,x_\perp) & = &
      \int {{d^2k_\perp} \over {(2\pi)^2}}
       {1 \over \sqrt{2\omega}}
      \left\{ a_1^a(\vec{k}_\perp) {1 \over \tau^{3/2}}
      e^{ik_\perp\cdot x_\perp -i\omega \tau} + C. C.
        \right\} \nonumber \\
        \vec{\alpha}^{a,i} (\tau,x_\perp) & = &
	\int {{d^2k_\perp} \over {(2\pi)^2}}
        \kappa^i
{1 \over \sqrt{2\omega}} \left\{ a_2^a(k_\perp)
{1 \over \tau^{1/2}}
      e^{ik_\perp x_\perp-i\omega \tau} + C. C. \right\}
\end{eqnarray}
In this equation, the frequency $\omega = \mid k_\perp \mid$, and the
vector
\begin{eqnarray}
	\kappa^i = \epsilon^{ij} k^j/\omega
\end{eqnarray}
The notation $+ C. C.$ means to add in the complex conjugate piece.

With the above form for the fields, the expressions for the number and energy
densities of the partons is straightforward. For late times, near $z=0$, the
they look like~\cite{blaimuell}
\begin{eqnarray}
	dE = {{dz} \over t} \int {{d^2k_\perp}\over {(2\pi)^2}}
	\omega \sum_{i,b} \mid a_i^b(k_\perp) \mid^2
\end{eqnarray}
Recalling that $dy = dz/t$, we find that
\begin{eqnarray} {{dE} \over {dyd^2k_\perp}} =
{1 \over {(2\pi)^3}} \omega \sum_{i,b} \mid a_i^b(k_\perp) \mid^2
\end{eqnarray}
and the multiplicity distribution of gluons is
\begin{eqnarray} {{dN} \over
{dyd^2k_\perp}} = {1 \over \omega} {{dE} \over {dyd^2k_\perp}}
\end{eqnarray}
The characteristic time scale for the dissipation of the non-linearities in the
equations for the time dependent Weizs\"acker-Williams fields can be estimated
from dimensional arguments. This is of the order $\tau = 1/p_\perp\equiv 1/
\alpha_S\mu$. Note that for large $\mu$'s, i.e., large parton densities, this
characteristic time gets smaller as one may intuitively expect.

\subsection{Onium--onium scattering}
\vspace*{0.4cm}

An interesting re--formulation of the low $x$ problem has been developed
recently in a paper by  A.~H.~Mueller~\cite{al1}. He considers an ``Onium"
(heavy quark--anti-quark) state of mass $M$ for which $\alpha_S(M)\ll 1$. In
weak coupling, the $n$-- gluon component of the onium wavefunction obeys an
integral equation whose kernel in the leading logarithmic and large $N_c$ limit
is precisely the BFKL kernel we discussed in the introduction. The derivation
relies on a picture in which the onium state produces a cascade of soft gluons
strongly ordered in their longitudinal momentum; the $i$--th emitted gluon has
a longitudinal momentum much smaller than the $i-1$--th.

The procedure to compute the wave function can be recast in terms of
hamiltonian perturbation theory in the following way: The Light Cone QCD
Hamiltonian~\cite{LV1} can be split into two pieces: $P^- = P^{-}_{0} +
P^{-}_{int}$, where $P^{-}_{0}$ and  $P^{-}_{int}$ are the non--interacting and
interacting pieces. Detailed expressions are given in Ref.~(15).

In the Infinite Momentum frame, where the parton picture can be applied, the
interaction Hamiltonian acting on an onium state produces an state with a
quark--antiquark pair and an extra gluon. First order perturbation theory can
be used to give an expression for that state according to
\be
   |\psi^{(1)}>=\sum_{q_1,q_2}\frac{<q_1,q_2,p-q_1-q_2\left|P^{-}_{int}\right|
   k_1,p-k_1>}{E^{(0)}_{k_1}(2-part) - E^{(0)}_{q_1,q_2}(3-part)}
   |q_1,q_2,p-q_1-q_2>\label{eq:nextstate}
\ee
where $E^{(0)}_{k_1}(n-part)$ is the eigenvalue of $P^{-}_{0}$ on a state
consisting of n free (quasireal) particles on mass shell;  $k_1$,  $p-k_1$ are
the quark and antiquark momenta and the sum is over all $q_1$, $q_2$ consistent
with overall momentum conservation. Notice that the energy denominator in
equation~(\ref{eq:nextstate}) will be equal to the (Light Cone) energy of the
created gluon on mass shell, namely $k_{1t}^{2}/2k_{1}^{+}$, due to the strong
ordering assumption. The state with n gluons and/or m quarks (anti--quarks) can
be obtained from the vacuum by applying n and/or m times the corresponding
creation operators appearing in the Light Cone definition of the fields in the
usual way,
\be
   A_{i}^{\alpha}(x) & = & \int \frac{d^3k}{(2\pi)^3\sqrt{2 k^+}}
                           \sum_{\lambda =1}^{2}
   \epsilon_{i}(\lambda , k) \left[ a^{\alpha}(\lambda , k)e^{ikx} +
                                    a^{\dagger \alpha}(\lambda , k)e^{-ikx}
                                                    \right] \nonumber \\
   \Psi(x) & = & \int \frac{d^3k}{(2\pi)^3}\left[b(k)e^{ikx} +
                 d^{\dagger}(k)e^{-ikx}\right].
   \label{eq:gluonquarkfields}
\ee
The wave function in momentum space is obtained by projecting the
state~(\ref{eq:nextstate}) onto the particular set of particle momenta. The
gluon number, is obtained by squaring the wave function and adding its terms
coherently. States with more than one gluon can be created by successively
applying the same procedure to the already created states, and the gluon
density to that order can be computed by squaring coherently the wave function
so obtained.

In the large $N_c$ limit, the $n$ gluons can be represented as a collection of
$n$--dipoles. Hence, in high energy onium--onium scattering, the cross section
is proportional to the product of the number of dipoles in each onium state
times the dipole--dipole scattering cross section~\cite{al2}. This cross
section is given by two gluon exchange~\cite{BFKL} (the pomeron). More
complicated exchanges are also possible.

It will be interesting to see if this approach can be extended to the problem
of nuclear scattering--how does it relate to the approach in the first part of
this section? That there are parallels between the two approaches is
evident--even though the large $N_c$ limit is a great simplification in the
onium case.

\section{Summary and outlook}

In this work, we have discussed a many body approach to computing parton
distribution functions for large nuclei for $x\ll A^{-1/3}$. Knowledge of these
structure functions is crucial in formulating the initial conditions for
ultrarelativistic nuclear collisions. Furthermore, such a first principles
calculation would eliminate the large uncertainities (especially at LHC) in jet
cross sections and like rates~\cite{Kari}. One may also apply these methods to
understand the systematics of nuclear shadowing in deeply inelastic scattering
experiments. Much work needs to be done in these directions.

There are several open theoretical questions which need to be addressed. A
pressing question is with regard to the applicability of weak coupling methods
to large nuclei. Is $\mu^2\sim A^{1/3}$ large enough? For the largest nuclei,
we have at most $\mu^2\sim 2$ GeV$^2$. At first blush, this seems extremely
unlikely. However, due to the possibility of a hierarchy of scales in the
problem (in analogy to finite temperature QCD), such a conclusion may be unduly
pessimistic. {\it If weak coupling methods are not applicable for computing the
parton distributions of the largest nuclei, it is unlikely that they will ever
apply during the subsequent stages of the evolution}. This would imply that
a weakly interacting Quark Gluon Plasma would never be formed!

Another question we would like to address is whether the methods discussed for
large nuclei can be extended to hadronic collisions. This depends on whether we
can find a way to perform the quantum mechanical sum over the external sources
in the partition function in Eq.~(\ref{qmsum}).

The $x\rightarrow 0$ limit of QCD is a very interesting one because many
features of the theory simplify in this limit. For instance, it has been shown
recently that this limit, for $N_c\rightarrow \infty$ is an exactly solvable
theory~\cite{korchemsky}. It will interesting to see whether these results can
be recovered in the many body formalism discussed in this paper.

\section{Acknowledgements}
This talk summarizes work begun with Larry McLerran and continued with Larry
and his students: Alejandro Ayala and Jamal Jalilian--Marian. I thank them all
for teaching me much of this material. I have benefitted from discussions with
Dietrich B\"{o}deker, Patrick Huet, Yuval Kluger, Sasha Makhlin, Berndt
M\"{u}ller, Paolo Provero and Heribert Weigert. I also wish to thank the INT
faculty for their moral support. This work was supported by the Department of
Energy under Grant No. DE--FG06--90ER--40561 at the Institute for Nuclear
Theory.

\end{document}